\documentclass[authorversion]{acmart}
\AtBeginDocument{%
  \providecommand\BibTeX{{%
    \normalfont B\kern-0.5em{\scshape i\kern-0.25em b}\kern-0.8em\TeX}}}

\usepackage{multirow}
\usepackage{enumitem}

\copyrightyear{2024}
\acmYear{2024}
\setcopyright{acmlicensed}\acmConference[LAK '24]{The 14th Learning Analytics and Knowledge Conference}{March 18--22, 2024}{Kyoto, Japan}
\acmBooktitle{The 14th Learning Analytics and Knowledge Conference (LAK '24), March 18--22, 2024, Kyoto, Japan}
\acmPrice{15.00}
\acmDOI{10.1145/3636555.3636906}
\acmISBN{979-8-4007-1618-8/24/03}

\begin{document}

\title[Temporal and Between-Group Variability in College Dropout Prediction]{Temporal and Between-Group Variability in College Dropout Prediction}

\author{Dominik Glandorf}
\email{dominik.glandorf@student.uni-tuebingen.de}
\affiliation{%
  \institution{University of Tübingen}
  \country{Germany}
}

\author{Hye Rin Lee}
\email{hyerin@udel.edu}
\affiliation{%
  \institution{University of Delaware}
  \country{USA}
}

\author{Gabe Avakian Orona}
\email{gabriel.orona@uni-tuebingen.de}
\affiliation{%
  \institution{University of Tübingen}
  \country{Germany}
}

\author{Marina Pumptow}
\email{marina.pumptow@med.uni-tuebingen.de}
\affiliation{%
  \institution{University Hospital Tübingen}
  \country{Germany}
}

\author{Renzhe Yu}
\authornote{Equal contribution}
\email{renzheyu@tc.columbia.edu}
\affiliation{%
  \institution{Teachers College, Columbia University}
  \country{USA}
}

\author{Christian Fischer}
\authornotemark[1]
\email{christian.fischer@uni-tuebingen.de}
\affiliation{%
  \institution{University of Tübingen}
  \country{Germany}
}

\renewcommand{\shortauthors}{Glandorf et al.}

\begin{abstract}
Large-scale administrative data is a common input in early warning systems for college dropout in higher education. Still, the terminology and methodology vary significantly across existing studies, and the implications of different modeling decisions are not fully understood. This study provides a systematic evaluation of contributing factors and predictive performance of machine learning models over time and across different student groups. Drawing on twelve years of administrative data at a large public university in the US, we find that dropout prediction at the end of the second year has a 20\% higher AUC than at the time of enrollment in a Random Forest model. Also, most predictive factors at the time of enrollment, including demographics and high school performance, are quickly superseded in predictive importance by college performance and in later stages by enrollment behavior. Regarding variability across student groups, college GPA has more predictive value for students from traditionally disadvantaged backgrounds than their peers. These results can help researchers and administrators understand the comparative value of different data sources when building early warning systems and optimizing decisions under specific policy goals.
\end{abstract}



\maketitle

\section{Introduction}
Preventing college dropout is a long-lasting goal of modern post-secondary education institutions \cite{fagioli2020role}. Succeeding Tinto's groundbreaking theory of academic integration \cite{tinto1973dropout}, large-scale administrative data and machine learning (ML) algorithms have been leveraged to build early warning systems (EWS) for student attrition in the recent decade~\cite{Fischer.2020}. However, the definition and nature of college dropout can vary significantly across institutional contexts, student populations, and application scenarios. Therefore, the full potential of early warning algorithms has yet to be systematically evaluated. Modeling decisions regarding the time of prediction and potentially different dropout mechanisms across student subgroups have to be better understood to build robust and reliable prediction systems.

This study aims to bridge the understanding of the complex dynamics of student dropout factors and real-world applications of college dropout models. Our analyses focus on the relevance of individual predictors and potential group differences, such as gender or other traditionally underrepresented groups, for dropout risks. We integrate the temporal dimension of the dropout prediction problem by comparing different time points at which dropout can be predicted. We target both further hypothesis-driven research and the construction of early warning systems (EWS) by answering the following research questions (RQs):

\begin{enumerate}[label=RQ\arabic*:]
    \item How well do college dropout prediction models perform when utilizing only administrative data, and which predictors within these models are the most important?
    \item How do the predictability and relevance of predictors of college dropout temporally change throughout enrollment?
    \item How do the predictability and relevance of predictors of college dropout vary between different student populations (i.e., underrepresented minorities, low-income-family, female, STEM, first-generation college students)?
\end{enumerate}

\section{Related work}
\subsection{Definition of College Dropout}
\label{sec:definition}

The opposite of academic success, i.e., no graduation or the absence of course grades in a given period, generally defines dropout \cite{Aulck.2019, Hellas.2018}. Academic success has been measured by time to degree, time of absence, or graduation \cite{Berens.2018, Casanova.2018}. Therefore, the following dimensions appear most helpful in defining dropout: retention, non-completion of the degree program, and elapsed time. Retention refers to whether a student reports back at the beginning of the term \cite{Aulck.2019}. Non-completion refers to students not returning and at the same time not having completed the program \cite{Bird.2021}. Elapsed time refers to the number of consecutive terms students were not enrolled in coursework \cite{solis2018perspectives}. Using these principles, we consider "a student that has not taken any courses at the university for at least four consecutive terms and has not completed the degree program” as dropout.

\subsection{Predictors of College Dropout}

\subsubsection{Administrative Data}
This data source refers to the information collected by the institution at the onset of and throughout a student’s college trajectory \cite{Fischer.2020}. It includes variables such as demographics (e.g., ethnicity, income status, first-generation status), academic performance measures (e.g., high school grade point average, scholastic aptitude test scores), and course-level outcome data (e.g., final grades). High school grade point average (GPA) and entrance test scores are mainstay predictors in most educational studies investigating college dropout \cite{Aulck.2019,Beaulac.2019, Berens.2018, DelBonifro.2020, Hellas.2018, Kemper.2020, Mason.2018, OrtizLozano.2020}. Similarly, demographic characteristics such as age, ethnicity, gender, and socio-economic status predict student dropout \cite{Aulck.2019,Berens.2018,DelBonifro.2020, Gray.2016,Hellas.2018, Kemper.2020,kotsiantis2003preventing, Mason.2018}. However, some studies indicate significant contributions to dropout predictions \cite{Berens.2018, Hellas.2018, Kemper.2020, Kovacic.2010, Reisel.2010}, whereas other studies fail to replicate such findings \cite{Herzog.2005, Mason.2018, Yu.2021}. More work is needed to understand how various demographic characteristics predict students’ academic success. 

\subsubsection{Alternative predictors}
Surveys add various motivational belief constructs, such as academic self-efficacy, values, and motivation, to the traditional predictors \cite{DiazMujica.2019, Gray.2016, Hellas.2018, Jeno.2018, DesJardins.2002}. However, the degree to which these attributes improve dropout predictions is debated, with modest improvements reported in prior work \cite{fagioli2020role}. The cost of this type of data and low response rates to non-obligatory student surveys (e.g., 9\% in \cite{saele2023note}) led us to omit survey data in our prediction models. A growing set of predictors is derived from learning management systems to predict students’ engagement \cite{bernacki2020predicting,hong2020latent}. Clickstream data refers to time-stamped records of student interactions triggered by the use of digital course material \cite{phippen2004practical,roll2015understanding}. Clickstream-based measures, such as idle time, number of keystrokes, and frequency of clicks within a particular page of the online learning environment indicate students’ engagement \cite{Fahd.2022}. Unfortunately, the heterogeneity of these environments and course-dependent use cases make this type of data hardly available for administrators as an off-the-shelf predictor in ready-to-scale prediction models.

\subsection{Prior early warning systems}
In recent years, the described data sources have been used for EWSs to improve educational practice. Early examples of such uses include Arnold and Pistilli’s study \cite{arnold2012course}, which leveraged demographic, learning-management-system, and previous academic records to identify students at risk of not being retained in courses. Similarly, Brown et al. \cite{brown2016and} utilized standardized test scores, course information, and demographics to implement early warnings for the performance of students enrolled in general education programs. More recent studies have branched towards additional data sources. A model that added formative assessments and online activity to predict final grade outcomes supplied approximately a 94\% accuracy rate by week 6 \cite{howard2018contrasting}. More recently, used E-book log data and Wifi connections served to predict the risk of course failure \cite{akccapinar2019using, zhao2022identifying}. Our approach has the potential to achieve similar predictiveness by incorporating only a cheap subset of these rich data sources. It can create a non-perfect but easy-to-implement first-level detection system that can indicate which at-risk students need more careful examination.

\subsection{Temporal dimension and subpopulations}
\label{sec:temporal_lit}
Dropout is often predicted only at one single point, such as the time of initial enrollment \cite{Aulck.2019}, end of the first term \cite{matz2023using, Dekker.2009}, or end of the first year \cite{Aulck.2019, Beaulac.2019}. Different prediction time points were typically only spread across one term to make course-related predictions \cite{akccapinar2019using, OrtizLozano.2020, zhao2022identifying}. Only a few studies predict at times ranging from initial enrollment until the end of the second year \cite{Hartl2019_1000096613}, and a single one reported the change of predictor importance for one of their models \cite{Berens.2018}. Apart from that, changes in college dropout factors have been only tracked over cohorts \cite{tanvir2021exploring}. This change of predictor importance between two or more time points within a student's trajectory was only systematically tracked for high school dropout \cite{franklin2016comparing} or within a survival analysis of college dropout focused on the time point of dropout \cite{ishitani2002longitudinal}. Therefore, we emphasize a systematic comparison between time points in this study concerning the prediction quality and data sources.

Group-specific dropout factors were previously only modeled in classical statistical models (regressions and structural equation modeling) \cite{suh2007predictors, archambault2017individual}. To the best of our knowledge, no prior work using ML methods has focused on group-specific college dropout predictors. We could identify almost no studies that reported the change in importance of these predictors on the between-group dimension. Only one study reported how factors may vary between public and private institutions \cite{Berens.2018}. We see this as a chance to combine the strength of ML methods with traditional interaction analyses (see Section \ref{sec:design}).

\section{Method}
\subsection{Study Setting and Sample}
This study is conducted at the University of California, Irvine, a large public research university in Southern California, that enrolls more than 25,000 undergraduate students. This university features a diverse undergraduate student body and received federal designations as a Hispanic-Serving Institution (HSI) and an Asian American and Native American Pacific Islander-Serving Institution (AANAPISI). Data for this study were provided from a multitude of services that collect and curate institutional data, including Admissions, the Registrar's Office, the Office of Institutional Research, the Office of Information Technology, and the Office of Financial Aid and Scholarships. Notably, this study is tied to a large institution-wide measurement project to understand the value of undergraduate educational experiences and promote evidence-based models of undergraduate student success trajectories \cite{arum2021framework}. The investigation uses data from six cohorts of degree-seeking non-transfer students (2011-2016 entrance dates) to capture both four- and six-year graduation rates. This led to a total sample size of 33,133 students, with records of 367,761 terms and 1,466,260 course enrollments, spanning twelve years of data. 

\subsection{Research Design} %
\label{sec:design}
To answer RQ1 and identify the best model, we evaluate all models' ability to predict dropout after one and two years of the initial enrollment. Based on typical dropout, we select two observation spans to account for dynamic relations between predictors and dropouts over time. RQ2 and RQ3 are computed based on the best-performing model from RQ1.

RQ2 analyzes the temporal dynamics of the dropout process. Ten subsets of the data span periods of up to the first three years of study (three terms per year). Starting from information available at the moment of the first enrollment, data obtained later than $n$ \textit{terms} after enrollment is discarded for $n\in \{0,1,...,9\}$. Students already known to have dropped out are excluded from later subsets. For each observation span, the relative importance of predictors is identified to track their changes over time. For time points later than this, models were not analyzed anymore due to the very low base rate of dropout by then.

RQ3 is structurally analogous to RQ2 but aims to identify differences between subgroups of students. We choose the attributes \textit{female}, \textit{first-generation student}, \textit{low-income family}, \textit{underrepresented minority}, and \textit{STEM major} (students in science, technology, engineering, and mathematics degrees) for the comparison of predictability and predictor relevance as they are often of particular interest for college administrators. The main analysis is conducted for a respective observation span of three terms, but a robustness analysis has been carried out for observation spans of two and four terms (see supplementary material\footnote{\url{https://github.com/dominikglandorf/college-dropout-prediction}}).

\subsection{Data}

\subsubsection{Outcome}
Dropouts are identified according to our dropout definition (see Section \ref{sec:definition}) and marked as such if students show at least four consecutive terms of no re-enrollment and were never reported to graduate. This leads to a prospective dropout rate of 13.2\% after the first year and 11.4\% after the second year. The descriptive statistics of all predictors and the conditional dropout rates can be found in the supplementary material.

\subsubsection{Pre-entry predictors}
Demographic data is usually available at the time of admission and remains invariant. Gender is simplified to a binary variable of whether a student is \textit{female}. The \textit{age at enrollment} is derived from the date of birth. \textit{International students} are annotated with the additional information if they took the TOEFL. The \textit{ethnicity} is captured as "Asian / Asian American", "Black", "Hispanic", "Indigenous", or "White non-Hispanic". For RQ3, "Black", "Hispanic", and "Indigenous" are summarized to the binary label \textit{underrepresented minority}, following a standard definition \cite{estrada2016}. The \textit{citizenship} is indicated as "US Citizen", "Permanent Resident", and "Not US Citizen". On the geographic scale, the \textit{residency within the state} at the time of application is known as "In-State", "Bona Fide", and "Out-of-State". The \textit{geographical category} contains more specific categorical information about residency before enrollment: "Foreign Country", "Out-of-State", "Northern California", "Southern California", and "University County". The university's \textit{distance from home} is enquired at the same point. Students within the \textit{first generation} of their family to study and students from a \textit{low income} family are flagged. The \textit{parents' education} is indicated per parent with the categories "No high school", "Some high school", "High school graduate", "Some college", "2 year college grad", "4 year college grad", and "Postgraduate study". The \textit{household size} at the time of admission is registered as the number of members, capped at a maximum of six. A binary variable indicates whether a student is a \textit{single parent}. Lastly, it is stated if a student is an \textit{English language learner} (i.e., non-native speaker).

Performance data collected prior to the studies contain the \textit{high school GPA} as well as the \textit{math}, \textit{writing}, and \textit{reading entry test score}. The \textit{best score in Advanced Placement (AP) exams} is used when available; otherwise, it is set to 0. The resulting year of study in the first term ("Freshman", "Sophomore", "Junior/Senior") is also recorded.

\subsubsection{Post-entry predictors}
Stemming from term-level information, the data contains the current \textit{number of declared majors} with the corresponding \textit{number of school affiliations}. The \textit{primarily affiliated school} and \textit{major} were indicated as a categorical variable. The number of \textit{changes of major}, \textit{school}, and \textit{total enrolled terms} are respectively derived from multiple term records. Note that our definition of the outcome only considers students as dropouts as soon as they have not re-enrolled for four consecutive terms. A binary flag indicates whether a student was declared as \textit{honors} for at least three terms and if at least one of their majors is a \textit{STEM major}. Predictors also include the average \textit{number of courses} taken per term and the \textit{current year of study}.

On the course level, both demographic and performance data are captured. The \textit{number of credits}, if the course has been \textit{passed}, and the numeric \textit{final grade} indicate the performance. The \textit{number of total students} and the relative amount of students of the \textit{same gender}, \textit{first-generation status}, and \textit{ethnicity} are captured as demographic indicators. All numeric information is first aggregated per term and then incorporated into cumulative averages up to the time point of prediction. The \textit{linear change in the number of credits} from the first to the current term is also calculated. Another statistic is the \textit{number of credits relative to the major's average}, and whether taken courses were \textit{offered by a school of one of the majors}.

\subsection{Modeling}
\label{sec:models}
A range of binary classification models is used in dropout prediction tasks \cite{shafiq2022student, akccapinar2019using, Gray.2016, Mason.2018, delen2010comparative, Aulck.2019, Berens.2018, Bird.2021, Wagner.2020}. We trained them on all predictors except the declared major as it shows redundancy with the school variable and would introduce too many variables. The \textit{logistic regression}, which assumes a linear relation between predictors and logarithmic odds of the outcome, is widely used. Besides being relatively simple to apply, it usually provides accurate predictions \cite{Fahd.2022}. The used ML methods include \textit{random forests} (RF), which are based on decision trees. Each tree recursively splits the data into two subsets based on the feature out of a random feature selection of a specified size that yields the best class impurity until the tree is grown to a specified size. An RF is assembled of a specified number of decision trees trained on different subsets of the training set. The predictions are averaged to make a robust prediction that maintains the quality of the individual trees. We use the R implementation in the package randomForest \cite{randomForest}.

The \textit{support vector machine} (SVM) chooses the position of a hyperplane in a multidimensional space, optimizing the class separation and the margin to their data points, incurring a certain cost on violations. Non-linear kernels that transform the input data into high-dimensional spaces often perform best. Performance also depends on regularization for the decision boundary (cost) and class weights. In the case of radial basis functions as a kernel, one must additionally specify the radius of influence (gamma). We use the implementation in the R package e1071 \cite{e1071}. The \textit{naive Bayes} classifier assumes an independent effect of categorical predictors on the outcome. Hence, it predicts its joint probability based on the observed class frequencies. Therefore, continuous predictors are discretized. The classifier can integrate a regularization value to generalize better to joint probabilities unobserved in the training data (Laplace parameter). We use the implementation in the R package e1071 \cite{e1071}.

\textit{k-nearest neighbors} identifies the k closest instances in terms of the Euclidean distance and predicts class membership based on the majority class within the neighborhood. Categorical predictors have to be dummy-coded for this purpose. We use the implementation in the R package class \cite{class}. Feed-forward \textit{neural networks} model non-linear functions by hierarchically applying linear transformations and non-linear activation functions to the input predictors. The two output neurons representing the two classes are normalized to probabilities using the softmax function. The error function is the binary cross-entropy loss, which is used to train the model weights and biases via backpropagation. We use the R interfaces to Keras \cite{keras} and TensorFlow \cite{tensorflow} using Python 3.10.

\subsubsection{Missing data imputation and hyperparameter tuning}
\label{sec:missing_data}
\begin{table}
  \caption{Sets of hyperparameter values used in grid search tuning procedure by model. |predictors|: total number of predictors}
  \label{tab:hyperparameters}
  \begin{tabular}{lll}
    \toprule
Model & Hyperparameter & Values \\
\midrule
Logistic regression & -    &   -     
\\
\midrule
k-Nearest Neighbors     & Number of nearest neighbors                & 9, 19, 39, 59, 99, 199, 299        \\
\midrule
Naive Bayes             & Laplace regularization                        & 0, 0.1, 0.5, 1.0                     \\ \midrule
\multirow{4}{*}{Neural Network}          & Number of neurons in first layer               & 256, 512, 1024                     \\
                        & Neurons in second layer (ratio of first)       & 0\%, 25\%, 50\%                       \\ 
                        & Dropout rate after each layer & 0\%, 50\%                             \\ 
                        & Training epochs                            & 5, 10                              \\ \midrule
\multirow{2}{*}{Random Forest}          & Number of grown trees                      & 500, 1000, 1500                    \\ 
                        & Number of candidates for split    & 3, 5, 6, 7                         \\ \midrule
\multirow{4}{*}{Support Vector Machine} & Kernel                                     & RBF, linear, polynomial \\ 
                        & Cost                                       & 0.1, 0.5, 1.0                        \\
                        & Gamma (for RBF kernel)                     & \{0.01, 0.1, 1.0, 10.0\}/|predictors|   \\ 
                        & Class weight for dropouts                  & 1, 3, 5\\
\bottomrule
\end{tabular}
\end{table}
Many predictive models require complete data. Due to the amount of missing data in some predictors, we prefer data imputation over keeping only complete data points. Creating multiple imputations to reflect the uncertainty in the missing data prediction and calculating results for all of them is a common approach. The R package mice \cite{mice} starts from simple baseline imputations and recursively repeats more sophisticated model-based predictions to improve them. We choose RF as the underlying single imputation method because of its suitability for categorical and continuous predictors. We generate ten imputed datasets, on which we run the entire training routine to ensure the robustness of our results against the randomness of the imputation. To ensure that classifiers perform best, their non-trainable parameters are optimized heuristically. This study uses grid search over all combinations of a careful selection of hyperparameter values (see Table \ref{tab:hyperparameters}). Our performance evaluation is based on the hyperparameter-tuned models and averaged metrics across imputations.

\subsection{Evaluation}

\subsubsection{Performance}
\label{sec:performance}
The performance of dropout prediction is estimated via 3-fold cross-validation, with a held-out test dataset to evaluate future performance. In dropout prediction, the imbalanced classes are often not addressed \cite{shafiq2022student}, e.g., when {\it accuracy} (ratio of correct predictions) is the only measure of performance \cite{OrtizLozano.2020, Beaulac.2019}. Overly predicting no dropout inflates accuracy without addressing the problem. Reporting per-class accuracy fixes the problem but hampers comparing datasets with different base rates. However, we include accuracy in the results to show how unsuitable the metric is in practice. It is calculated for 200 thresholds, ranging from 0 to 1, to choose the threshold that yields the maximum possible score optimally.

The more comprehensive receiver-operator characteristic (ROC) summarizes all possible thresholds for the \textit{sensitivity} (ratio of actual dropouts detected, also called \textit{recall}) and \textit{specificity} ($1 - \text{false positive rate}$) and can be summarized in a scalar, called area under ROC curve (AUROC), which is often used in dropout prediction. The precision-recall curve (PRC) is more informative for imbalanced classes because it incorporates the \textit{precision} (correct ratio of predicted dropouts) instead of the specificity. It also can be summarized in the area under PRC (AUPRC) \cite{fu2019tuning}. For all metrics, the best possible classifier with no predictors implies the baseline. Given the base rate for dropout is $r_d$, the baseline accuracy in binary classification is $1-r_d$. In the case of AUROC, it is $0.5$; for AUPRC it is $r_d$.

\subsubsection{Predictor Importance}
To ensure the meaningfulness of our results to administrators, the global scores of predictor importance are of most interest. We choose a model-agnostic approach to calculate the significance of single predictors, such that scores are available independent of the model performance ranking and can theoretically be compared across models. The \textit{Permutation Feature Importance} (PFI) measures how much the test set performance of a model decreases when one variable is randomly permuted \cite{Beaulac.2019}. For RQ2, the PFI is based on the more sensitive AUPRC. In RQ3, the different base rates led us to use the AUROC with its constant baseline. By excluding predictors with a variance inflation factor (VIF) larger than 5, we ensure that predictors are independent enough for meaningful PFI scores. The PFI is averaged over test sets and imputations in the same way as the performance metrics.

\section{Results}

\begin{table}
\centering
\caption{Estimated performance metrics of fine-tuned models on the entire data respectively up to one year and two years after first enrollment.}
\label{tab:performance}
\begin{tabular}{rllllll}
  \toprule Time of prediction & \multicolumn{3}{c}{After first year} & \multicolumn{3}{c}{After second year}\\
   \midrule Model & AUPRC & AUROC & Accuracy & AUPRC & AUROC & Accuracy \\ 
  \midrule Baseline & 0.131 & 0.500 & 0.869 & 0.112 & 0.500 & 0.888 \\ 
  Logistic Regression & 0.553 & 0.816 & 0.898 & 0.712 & 0.899 & 0.938 \\ 
  k-Nearest Neighbors & 0.493 & 0.785 & 0.890 & 0.620 & 0.863 & 0.922 \\ 
  Naive Bayes & 0.501 & 0.789 & 0.890 & 0.680 & 0.885 & 0.928 \\ 
  Neural Network & 0.561 & 0.823 & 0.899 & 0.722 & 0.903 & 0.939 \\ 
  Random Forest & \textbf{0.577} & \textbf{0.827} & \textbf{0.902} & \textbf{0.735} & \textbf{0.908} & \textbf{0.940} \\ 
  Support Vector Machine & 0.556 & 0.821 & 0.898 & 0.718 & 0.903 & 0.938 \\ 
   \bottomrule \end{tabular}
\end{table}

\subsection{RQ1: Dropout predictability and predictor importance of different models}
All models' performance in dropout prediction on the general population is summarized in Table \ref{tab:performance}. The metrics are almost invariant across the ten missing data imputations and hence averaged, enabling us to compare differences between the models independent of the data imputation (see standard deviations in supplementary material). The relatively small differences in the accuracy metric result from the high baseline accuracy. As predicted, the AUPRC shows a larger variation than the AUROC, which empirically justifies its use. Especially the area-based metrics show that all models perform above the baseline. The RF model emerges as the best by dominating all metrics by a notable margin at both prediction times. The neural network does perform second-best, whereas the traditional logistic regression shows performance almost on par with the SVM. These four models will be referred to as the top 4.

We define the most important predictors of student dropout in terms of their PFI as the most insightful information for administrators. These predictor importances are listed in Table \ref{tab:importance}. The number of essential predictors is relatively sparse compared to the overall number of predictors (39). Depending on the model and time of prediction, only seven to ten predictors impact the AUPRC by more than 1\%.

When predicting dropout after a student's first year, performance indicators related to grades and passing are most important, along with continuous enrollment (i.e., a higher number of enrolled terms) and being on track in the current year of study in all portrayed models. The acquisition of English as a second language and other demographic factors such as individual and peer ethnicity also impact the prediction. Overall, recorded behavior within the first year at college seems to have more impact than demographic variables. This pattern does not only occur for random forests but across all well-performing models. Depending on the specific model, some rankings might change by one or two positions, but the pattern is comparable.

Remarkable differences emerge if we compare the importance between the two prediction time points. A fine-grained analysis of importance change over time is the subject of RQ2. Two years after initial enrollment, the number of enrolled terms is by far the best predictor and is preferred by every top-4 model. In contrast, the college GPA has lost its initial importance. By the ensemble of the top 4 models, passed courses are preferred over the college GPA as a predictor. The number of passed courses and year of study become much more important. The fact that English learners are less likely to drop out is reflected by all the models and maintains a PFI of more than 1\% among the pre-entry predictors. Also, at this point, the ranking changes slightly when considering different models but the magnitudes and relative importance of predictor pairs are model-independent. 

\begin{table}[]
\caption{The 15 most important predictors after one and two years measured by Permutation Feature Importance. RF: Random Forest. Top 4: Mean of Random Forest, Linear Regression, Support Vector Machines, and Neural Network.\\Left: after one year, Right: after two years.}
\label{tab:importance}
\centering
\begin{tabular}{rlll}
\toprule  & Predictor & RF & Top 4 \\ 
  \midrule 1 & College GPA & 23.0\% & 16.1\% \\ 
  2 & Number of enrolled terms & 15.0\% & 16.2\% \\ 
  3 & Passed courses & 7.4\% & 11.6\% \\ 
  4 & Number of school affiliations & 4.9\% & 3.6\% \\ 
  5 & English language learner & 3.5\% & 3.8\% \\ 
  6 & Current year of study & 2.8\% & 1.6\% \\ 
  7 & Number of declared majors & 2.6\% & 1.9\% \\ 
  8 & Primarily affiliated school & 2.0\% & 2.6\% \\ 
  9 & Ethnicity & 1.3\% & 2.6\% \\ 
  10 & High school GPA & 1.1\% & 0.5\% \\ 
  11 & Same ethnicity in courses & 0.9\% & 0.6\% \\ 
  12 & Residency at application & 0.7\% & 0.8\% \\ 
  13 & Distance to university & 0.7\% & 0.6\% \\ 
  14 & Pre-entry reading score & 0.5\% & 0.5\% \\ 
  15 & Courses per term & 0.5\% & 0.4\% \\ 
   \bottomrule \end{tabular}
\quad
\begin{tabular}{rlll}
 \toprule  & Predictor & RF & Top 4 \\ 
  \midrule 1 & Number of enrolled terms & 48.3\% & 54.1\% \\ 
  2 & College GPA & 8.5\% & 3.5\% \\ 
  3 & Passed courses & 4.8\% & 7.5\% \\ 
  4 & Current year of study & 4.1\% & 2.8\% \\ 
  5 & Number of school affiliations & 2.5\% & 1.3\% \\ 
  6 & English language learner & 2.1\% & 2.3\% \\ 
  7 & Primarily affiliated school & 1.0\% & 1.4\% \\ 
  8 & Number of declared majors & 0.9\% & 0.4\% \\ 
  9 & Distance to university & 0.5\% & 0.4\% \\ 
  10 & Same ethnicity in courses & 0.5\% & 0.3\% \\ 
  11 & Ethnicity & 0.4\% & 1.1\% \\ 
  12 & Courses per term & 0.4\% & 0.2\% \\ 
  13 & Pre-entry reading score & 0.3\% & 0.3\% \\ 
  14 & Female & 0.3\% & 0.5\% \\ 
  15 & High school GPA & 0.3\% & 0.1\% \\ 
   \bottomrule \end{tabular}
\end{table}

\subsection{RQ2: Dropout predictability and predictor importance over time}
\begin{figure}
\centering
\includegraphics[width=0.9\textwidth]{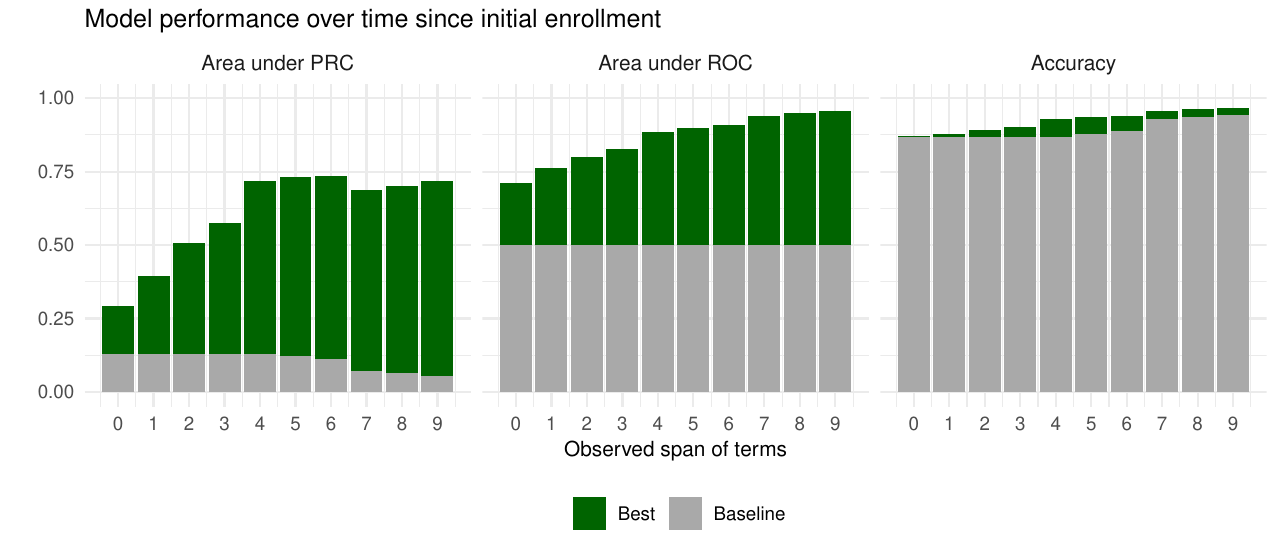}
\caption{Performance metrics for different time points of predictions against their respective baselines. PRC: precision-recall curve, ROC: receiver-operator curve. The baseline is a random prediction for curve-based metrics, while based on the best possible threshold for accuracy.}
\label{fig:RQ2_performance}
\end{figure}

The results for RQ2 and RQ3 are based on the RF because it performed best and due to comparable predictor rankings (and performance differences) between models. Figure \ref{fig:RQ2_performance} depicts the performance of this model trained with data up to a certain time point relative to the initial enrollment. The general trend is an increased model performance with a growing observation span. The fact that AUPRC is not monotonically increasing over time is due to the change in the base rate of dropout, which starts to shrink after five terms when the first students are known to have dropped out. The accuracy shows no improvement over the baseline at the pre-entry time point. Nevertheless, the two area-based metrics demonstrate that the model already outperforms the baseline at this early time point. Over time, and especially during the beginning of the second year,  accuracy stands slightly out from the baseline. The AUROC shows a steady increase, most strongly in the first terms. The most variation and increase can be observed in the AUPRC, which was expected to be the most suitable evaluation metric in this scenario.

\begin{figure}
\centering
\includegraphics[width=0.9\textwidth]{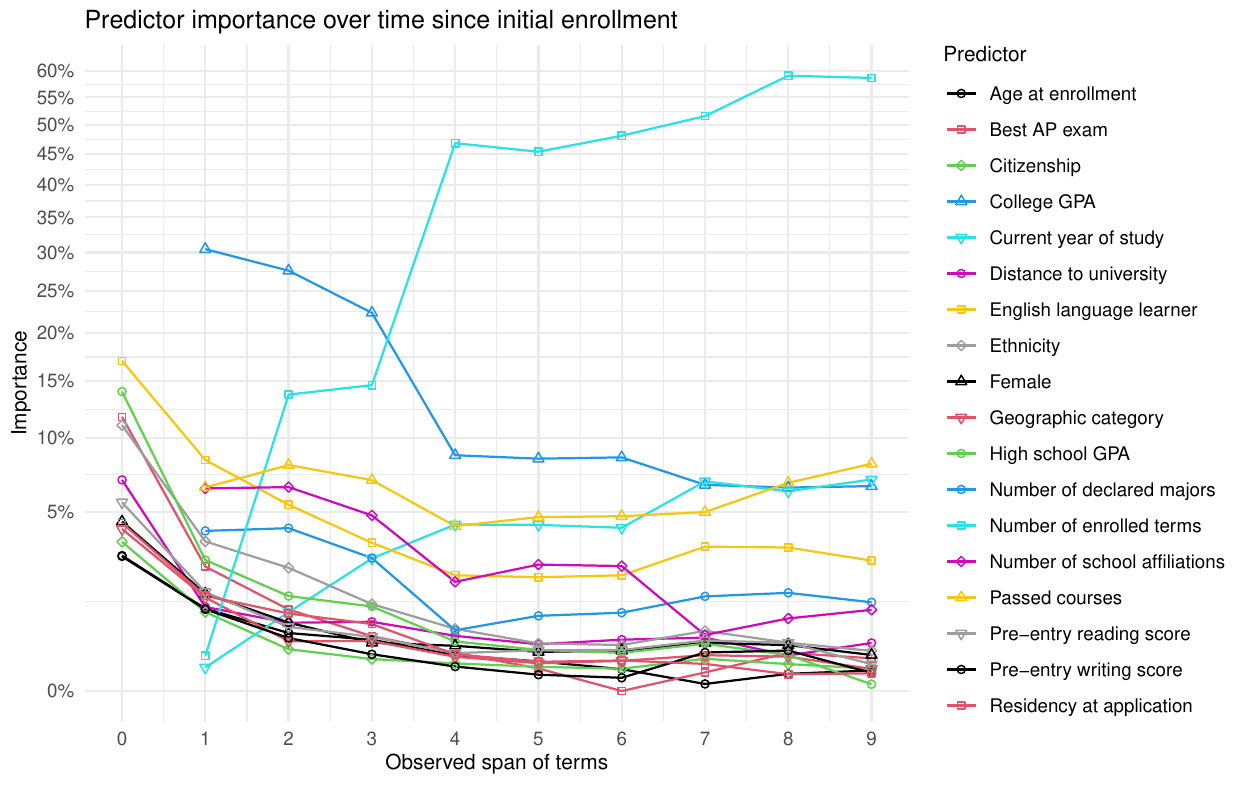}
\caption{Predictor importance for different time points of predictions. Predictors always below 2.5\% are omitted. Due to the root transformation of the importance for better readability, differences in the area below 5\% may seem aggravated.}
\label{fig:RQ2_importance}
\end{figure}

In Figure \ref{fig:RQ2_importance}, the change of relative predictor importance is plotted. At the time of initial enrollment, a mixture of demographic and performance indicators weighs the most in the prediction. The status as an English language learner contributes the most information, closely followed by high school GPA, the best AP exam score, and the student's ethnicity. The geographic information (mostly distance from home), the pre-entry reading score, and gender play minor but significant roles. All other predictors influence the performance by less than 4\%. The ranking is strongly perturbed as soon as behavioral data from college studies is available. By the end of first term, the GPA leads the ranking by a large margin (22.2\%). The English language learner status retains a part of the initial information, ranking second. The five most important factors after one term are completed by performance-related indicators: passed courses, number of school affiliations, and number of declared majors. Ethnicity follows with less than 4\% impact, which also applies to the high school GPA and best AP score by then. 

The relative importance ranking does not consolidate after the first term but continues to change. Most strikingly, the number of enrolled terms drastically increases its predictive importance over time. Variance in this predictor is possible from the second term on and directly reflects in an importance score of 13.7\% by then. At that time point, the average grade is still the most informative predictor (27.6\%). The relative ranking does not change much within the remainder of the first year of enrollment. After four terms, the number of enrolled terms eventually ranks first by a significant margin. The college GPA gets degraded to less than 10\%. English learner status still plays a role, although it is very minor. During the second year, the ranking stays rather stable. During the third year, the ratio of passed courses and the current year of study overtake the importance of GPA. The current year of study slightly gains importance, starting from the third term. Interestingly, the English language learner status gained some meaning during that period. 

\subsection{RQ3: Dropout predictability and predictor importance for subpopulations}

\begin{figure}
\centering
\includegraphics[width=1\textwidth]{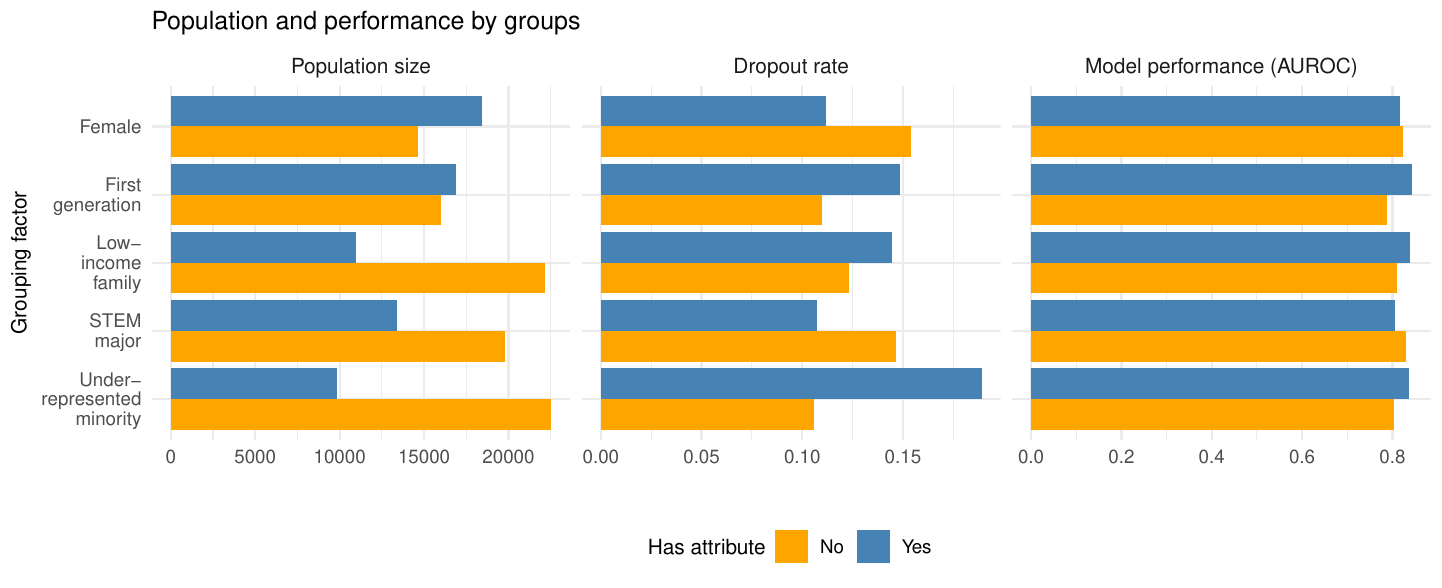}
\caption{Model performance and population sizes by grouping factors. Both are based on the data available one year after initial enrollment.}
\label{fig:RQ3_performance}
\end{figure}

To compare the dropout predictability and predictor importance between different groups of students, we resort to the AUROC due to its fixed baseline of 0.5 and its sufficient sensitivity to performance differences in RQ2, especially between the second and the fourth term. Figure \ref{fig:RQ3_performance} shows the predictive performance as a function of groups, along with the group sizes and dropout rates. Although there are some unbalanced grouping factors in terms of population size (e.g., only 33.1\% of students stem from low-income families) and dropout rates (18.9\% of students from an underrepresented minority drop out), the predictability of dropout does not vary enormously between the respective groups. Prediction seems slightly easier for first-generation students, non-STEM majors, and students from low-income families and underrepresented minorities (URM). 

\begin{figure}
\centering
\includegraphics[width=1\textwidth]{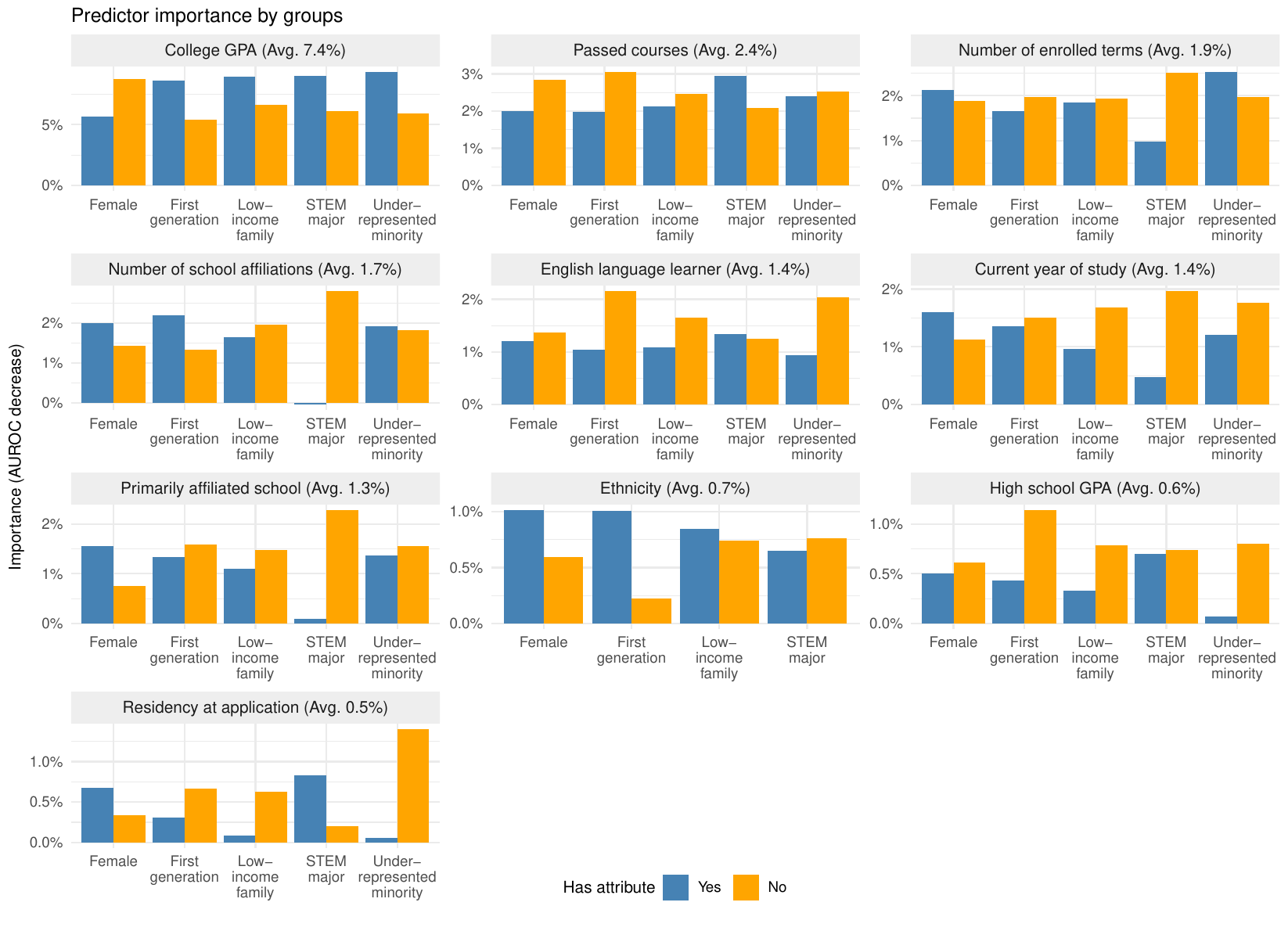}
\caption{Differences in predictor importance between groups when predicting dropout one year after initial enrollment. Twenty-nine predictors with a maximal score of 1\% or below for every group are omitted in this plot.}
\label{fig:RQ3_PFI}
\end{figure}

Figure \ref{fig:RQ3_PFI} visualizes the importance of predictors between the groups. The generally most important predictor after one year, college GPA, is the top predictor for all subpopulations. Despite its group-independent overall relevance, we can highlight some notable differences between groups. Within the PFI of the number of enrolled terms and passed courses ratio, we observe similar between-group differences that also rarely change the overall ranking of importance within one group. For generally less important factors, such as major and school information, current year of study, and English learner status, one can observe minor but potentially significant differences in absolute value.

For female students, the GPA is notably less relevant with respect to dropout than for non-female students which also applies to course passing. The number of enrolled terms contributes almost equal information to both group's predictions. Prediction for female students may benefit more from major and school information and the year of study. Ethnicity is also more relevant for female dropouts whereas English learner status does not vary in its importance between the two groups. First-generation student dropout stronger depends on GPA than its counterpart, for which course passing is more informative. The first-generation dropout prediction benefits more from the number of school affiliations and ethnicity, whereas English learner status and high school GPA show the opposite interaction effect. This trend is mirrored for students from low-income families. Only the number of school affiliations has a less pronounced effect. Moreover, the importance of ethnicity differs much less between the induced groups by this criterion.

Being part of a URM mostly means an increased importance of the same factors than for low-income families. However, English learner status is more important for non-URM students. We also observe the clear pattern that for URM students the high school GPA and residency location at the time of application are irrelevant while having predictive power for the remainder of students. Comparing STEM majors and non-STEM majors also reveals a prominent difference in GPA importance, being more predictive in the case of STEM majors. Course passing is also more helpful in predicting this group's dropout, whereas the number of enrolled terms is much more interesting for non-STEM majors. The same applies to the number of school affiliations, which are irrelevant for STEM dropout in this model. The pace of study is only relevant for non-STEM dropout.

\section{Discussion and Conclusion}

This study has successfully employed modern classification models to predict college dropout, relying on cost-effective large-scale administrative data that can be adopted for an early warning system. The model-agnostic PFI yielded valuable insights into the dropout factors. As novel contributions, we traced the predictability of dropout over a large span of the student lifecycle and the importance of single predictors. Moreover, the analyses address student heterogeneity by distinguishing the predictive importance of factors between important grouping factors of college success.

\subsection{Scholarly Significance}
Our model's performance is mostly comparable with previous studies using the same data type. Studies that made predictions after one term based on institutional data reported an AUROC between 0.69 and 0.88 (depending on the school) \cite{matz2023using} compared to our score of 0.76. In the literature, scores rise to the range between 0.81 and 0.93 after one year of enrollment \cite{Berens.2018, Aulck.2019}, where we achieve 0.83. Results for datasets from different institutions and regions remain hard to compare. However, considering the low overall impact of sociodemographic factors in our analyses and the standardized federal reporting of many predictors, the findings may be generalizable to other 4-year colleges in the US.

Due to the rich temporal structure of our data, we traced how dropout prediction factors evolve. Although Tinto's theory of dropout \cite{tinto1973dropout} emphasizes the longitudinal character of the dropout process, even recent studies did not include this critical dimension of dropout prediction (see Section \ref{sec:temporal_lit}). Therefore, we mapped how pre-entry information decreases in value over time and that college performance data is generally most valuable in the first year, whereas continued enrollment is most important after the first year. Academic integration (as measured by GPA) may become substituted by social integration (indicated by the number of enrolled terms) to predict dropout over time. Interestingly, prior work suggested the reverse \cite{ishitani2016time}. However, course peer composition measures were not highly predictive as potential indicators of social integration. An outstanding overall predictor was students' status as English language learners, which may be related to social integration and is worth investing more in-depth for future research. Overall, administrative data focuses on academic integration and contains limited information regarding social integration. 

The large-scale character of our data yielded large enough subpopulations to examine differences in predictive factors. We believe that this interaction analysis, based on ML models, can open up new directions in dropout prediction research. We found significant differences between the dropout factors for certain subgroups. Considering that college GPA, the ratio of passed courses, and the number of enrolled terms reflect a spectrum from performance to non-performance-related behavior, students from traditionally disadvantaged groups may be more reliant on grades regarding dropout. This could be an exemplified interaction of academic and social integration. It may be worth investigating if these groups are more likely to stay enrolled and perform worse instead of taking a semester off before dropping out. The most striking differences were found for students in STEM majors, whose dropout is better predicted by their grades, and course passing. Non-STEM dropouts are easier identified by the number of enrolled terms, number of majors, and school affiliations. Again, academic integration may be more crucial in STEM fields compared to other subjects for the student's commitment that eventually leads to dropout, resonating with the literature \cite{dika2016early}.

\subsection{Implications for Educational Stakeholders}
Our results further support the value of administrative data that every institution already has in standard formats \cite{Fischer.2020}, compared to costly interviews or cleaning-intensive process data. We provide our analysis scripts to encourage other universities to replicate the analyses with their data. Generally, our methodology allows administrators directly or by re-analysis to create a precise and smaller hence cheaper set of dropout predictors. First, the temporal dimension of the prediction can be considered when implementing an early warning system. At the moment of enrollment, predictions are much more error-prone than after the first year. Based on the change over time in dropout predictability, educational stakeholders may choose their individual preference along the trade-off between early interventions against dropout and accumulated evidence for actual dropout risk. Second, the selection of predictors to obtain can now be adapted to the actual time point where a prediction should be made. Although administrative data itself comes at a lower cost compared to other data sources, the integration of different information systems across a higher education institution always implies effort or may be subject to data privacy regulation \cite{Berens.2018}. As we have shown, pre-entry data rapidly loses value as soon as behavioral data at the college becomes available. Depending on the chosen time point for dropout prediction, administrators can estimate the value of predictors more precisely. Third, an EWS can be tailored to specific subpopulations of the student body using group-specific factors. We found that dropout prediction fortunately works almost equally on groups induced by various grouping factors. However, in the case of STEM majors, the predictor collection can vary as a function of the targeted group.

Overall, an effectively designed EWS may help reduce overall dropout via resource-efficient intervention by administrators. It allows for reducing false negatives by tolerating more false positives using the classification threshold. Nevertheless, the following risks should be mitigated: Biases in helpfulness associated with protected attributes \cite{Yu.2021}, which our results, fortunately, did not reveal; biasing faculty in their behavior towards at-risk students \cite{gentrup2020self}, which may be prevented sending warnings directly to students or to central consulting offices; and misuse of the EWS for admission decisions, which requires appropriate legislation.

\subsection{Limitations and Future Work}
When students at risk are identified, possible interventions may want to address underlying reasons. Most of the identified predictors in this study may serve as an explanation for the risk of dropout but are usually hard or even impossible to change directly. The prediction of dropout still requires careful human examination of individual cases.

Although we use the VIF to make sure that our univariate importance metric is meaningful, there is still an inherent correlation between most predictors. Differences in predictor importance are in this approach naturally dependent on the existence of the other predictors in the training. For example, while high school GPA loses its predictive value relative to the college GPA after the first term it does not mean that it becomes useless. Instead, other predictors may just be more suitable for subsequent performance. This possibility should be considered when deciding in favor or against the survey of certain predictors as an educational stakeholder.

The approach of this study hopes to identify some potential "blind spots" of traditional hypothesis-driven analyses. It provides more insights for theory development and further hypothesis testing compared to other more data- and algorithm-driven approaches that often underlie EWS used for predicting and reporting risks \cite{Jayaprakash.2014}. Similarly, our approach to using existing large-scale administrative data provides more cost-efficient alternatives to expensive questionnaire-based surveys or complex clickstream data to support educational administrations in their decision-making. Ultimately, we believe that the automatization of EWS may allow for adaptive student support systems to foster more learner-centered environments, enhance learning benefits, and reduce dropout risks \cite{Chatti.2019, Ifenthaler.2020}.

\bibliographystyle{ACM-Reference-Format}
\bibliography{EWS-bibliography}

\end{document}